\documentclass[twocolumn,tightenlines,floatfix]{revtex4}

\usepackage{epsfig} 
\usepackage{amssymb} 
\usepackage{amsmath} 
\usepackage{amsfonts} 

\newcommand{\be}{\begin{equation}}
\newcommand{\ee}{\end{equation}}
\newcommand{\ba}{\begin{eqnarray}}
\newcommand{\ea}{\end{eqnarray}}

\begin{document}

\title{Secondary atmospheric tau neutrino production}

\author{Alexander Bulmahn and Mary Hall Reno}
\affiliation{Department of Physics and Astronomy, University of Iowa, Iowa City, IA 52242}

\begin{abstract}
We evaluate the flux of tau neutrinos produced from the decay of pair produced taus from incident muons using a cascade equation analysis.  To solve the cascade equations, our numerical result for the tau production $Z$ moment is given.  Our results for the flux of tau neutrinos produced from incident muons are compared to the flux of tau neutrinos produced via oscillations and the direct prompt atmospheric tau neutrino flux.  Results are given for both downward and upward going neutrinos fluxes and higher zenith angles are discussed.   
We conclude that the direct prompt atmospheric tau neutrino flux
dominates these other atmospheric sources of tau neutrinos for neutrino
energies larger than a few TeV for upward fluxes, and over a wider range of energy for downward fluxes.
\end{abstract}

\maketitle

\section{Introduction}

Atmospheric muons and muon neutrinos dominate the atmospheric lepton flux for a lepton energy of 100 GeV \cite{bartol}.
For $E=100$ GeV, the dominant sources of atmospheric leptons are pion and kaon production by cosmic
ray interactions in the atmosphere (the ``conventional flux'')
\cite{lipari} at a height of 15-20 km \cite{pathlength}, 
followed by their decay. Cosmic ray production of
charm is suppressed at this energy, where the ``prompt flux'' of muon neutrinos is about three orders of magnitude
smaller than the conventional flux of muon neutrinos \cite{prs,tig}. Charm decays, via the $D_s$ meson, can also
produce tau neutrinos. The prompt tau neutrino flux is about one order of magnitude smaller than the prompt muon
neutrino flux \cite{Pasquali,stasto}.

Neutrino oscillations of atmospheric neutrinos can produce appreciable atmospheric tau neutrino fluxes in some energy ranges.
Indeed, the oscillations of upward muon neutrinos  to tau neutrinos 
as they propagate along the diameter of the Earth resulted in a convincing
deficit in muon neutrinos in the sub-GeV and multi-GeV event samples in the SuperKamiokande detector \cite{superk}.
Downward atmospheric neutrinos have a relatively short distance to propagate, so the oscillation probability is reduced.
At higher energies, neutrino oscillations are also suppressed.

Neutrino oscillations over astrophysical distances will provide for astrophysical tau neutrino fluxes. The neutrinos may come from pion and kaon production and decay in sources of cosmic rays, or they may come from more exotic sources, such as dark matter annihilation to neutrinos or particles which decay to neutrinos. 

In this Brief Report, we explore another source of tau neutrinos, namely tau pair production by atmospheric muons in transit through the Earth. These tau pairs may lose energy as they traverse a portion of the Earth, but for low enough energy, they do not since the tau lifetime is so short \cite{drss}. We provide here an evaluation of the tau neutrino flux from these tau pairs which we compare with the oscillated atmospheric and prompt atmospheric tau neutrino fluxes.  We find that the flux of tau neutrinos from muon production of tau pairs is higher than the oscillated tau neutrino flux in the downward direction for energies above $\sim 1000$ GeV.  In the upward direction, the oscillated flux is higher than that from muon production out to $\sim 10^5$ GeV.  In both cases, however, the tau neutrino flux from muon production of tau pairs is lower than the prompt atmospheric tau neutrino flux. 

In the next section we describe how the flux of tau neutrinos from tau pairs is evaluated. In Sec. III, we 
discuss the oscillated and prompt tau neutrino fluxes. Our results for the
downward and upward tau neutrino fluxes are shown in Sec. IV.

\section{Tau neutrinos from atmospheric muon tau pair production}

Muon production of electron-positron pairs as muons pass through materials is one of the dominant energy loss mechanisms for muons. While kinematically suppressed, tau pairs are also produced. Up to $E_\tau=10^6$ GeV and higher, tau energy
loss does not compete with tau decay \cite{drss}, with each decay resulting in a tau neutrino. While the tau pair production cross section is lower than the electron-positron pair production cross section by a factor of $\sim 10^{-7}-10^{-8}$ 
\cite{br1} at
1 TeV muon energies, there are orders of magnitude more muons than
tau neutrinos coming from direct production or from oscillations in
the downward direction.

To evaluate the tau neutrino flux from muon production of tau pairs, we use the $Z$-moment method often applied to atmospheric
neutrino calculation \cite{lipari}. The starting point is the coupled set of cascade equations in terms of differential lepton
fluxes $\phi_i$ which depend on the lepton energy and the column depth $X$,
\begin{eqnarray}
\frac{d\phi_\tau}{dX} &=& -\frac{\phi_\tau}{\lambda_{dec}}+S(\mu\to\tau)\\
\frac{d\phi_{\nu_\tau}}{dX} &=& S(\tau\to\nu_\tau) \ ,
\end{eqnarray}
where the source terms $S$ account for the production of taus or the tau decay,
\ba
S(\mu\to\tau) &=& \int_{E_\tau}^\infty dE_\mu\frac{N_A}{A} {\phi_\mu(E_\mu,X)}\frac{d\sigma_{pair}^{\mu\to\tau}(E_\mu,E_\tau)}{dE_\tau}\nonumber\\
&\equiv & \frac{\phi_\mu(E_\tau,X)}{\lambda_{pair}(E_\tau)}Z_{\mu\tau}(E_\tau)
\label{eq:zmutau} \\
S(\tau\to\nu_\tau)&=& \int_{E_{\nu_\tau}}^\infty dE_\tau \frac{1}{\rho} {\phi_\tau(E_\tau,X)}\frac{d\Gamma^{\tau\to\nu_\tau}(E_\tau,E_{\nu_\tau})}{dE_{\nu_\tau}}\nonumber\\
&\equiv& \frac{\phi_\tau(E_{\nu_\tau},X)}{\lambda_{dec}(E_{\nu_\tau})}Z_{\tau\nu_\tau}(E_{\nu_\tau})\ .
\ea
Here, we use the interaction length which depends on the tau pair production cross section
$\lambda_{pair}(E)=A/(N_A\sigma_{pair}(E))$ \cite{br1} and the decay length $\lambda_{dec}(E) = \rho E/(\Gamma_\tau\, m_{\tau})$
where $\Gamma_\tau=1/\tau_\tau$, the inverse lifetime of the tau lepton at rest, $N_A$ is Avogadro's number and $A$ is the nucleon number. 
The flux of muons depends on zenith angle $\theta$ as well, but since
we will look specifically at angles of $\theta=0,\ 180^\circ$, we will
not include the $\theta$ dependence explicitly. A detailed
parameterization of the $\theta$ dependence appears in, for example,
Ref. \cite{KBS}.

The differential energy distribution
of the tau production cross section 
for $Z_{\mu\tau}$ is evaluated numerically following Ref. \cite{br2}. A parameterization for the decay distribution in terms of $y = E_{\nu_\tau}/E_\tau$ for $Z_{\tau\nu_\tau}$ can be found in Ref. \cite{Pasquali}.

The atmospheric muon flux which appears in the $\mu\to \tau$ equation can also be represented by a transport
equation, but to first approximation, it can be written as
\be
\label{eq:mudepth}
\phi_\mu (E_\mu,X) = \phi_\mu(E_\mu^s,0)\exp(\beta X)\ .
\ee
In the above equation, $\phi_\mu(E_\mu^s,0)$ represents the muon flux at the surface of the Earth.
The conventional atmospheric muon flux from pion and kaon decay can be parameterized by \cite{KBS}
\ba
\label{eq:musur}
\phi_\mu^{\rm conv}(E_\mu^s,0) &=& 
\frac{0.175}{\rm GeV  cm^{2} s\,sr}\,\frac{1}{(E_\mu^{s}/{\rm GeV})^\gamma}\nonumber\\
&\times & 
\Bigl( \frac{1}{1+E_\mu^s/103\ {\rm GeV}}
\nonumber \\
&+& \frac{0.037}{1+E_\mu^s/810\ {\rm GeV} } \Bigr) \ .
\ea
Here $103$ GeV and $810$ GeV represent the pion and kaon critical energies, which separate the high and low energy contributions to the muon flux.  The spectral index  is $\gamma= 2.72$. 

We also include the muon flux from charm production and decay in the atmosphere,
the so-called ``prompt'' flux. Here, we use the flux
\ba
\phi_\mu^{\rm prompt}(E_\mu^s,0) &=& 
\frac{2.33\times 10^{-6}}{\rm GeV  cm^{2} s\,sr}\,\frac{1}{(E_\mu^s/{\rm GeV})^{\gamma'}}\nonumber\\
&\times & 
\frac{1}{1+E_\mu^s/3.08\times 10^6\ {\rm GeV} } 
\ea
from a dipole model evaluation of the charm production cross section \cite{enberg},
although other choices are possible \cite{prs,tig,stasto}. 
The spectral index for the prompt flux is $\gamma'=2.53$. Note that the prompt muon neutrino flux equals the prompt muon flux.

Taking energy losses through ionization and radiative processes into account, the muon energy at depth is related to the surface energy through
\be
\label{eq:esur}
E_\mu^s = \exp(\beta X)E_\mu+(\exp(\beta X)-1)\frac{\alpha}{\beta}\ 
\ee
in terms of the $\alpha=2.67 \times 10^{-3}\, {\rm GeV\,cm^2/g}$ and
$\beta = 2.4\times 10^{-6} \, {\rm cm^2/g}$ for $E_\mu<3.53\times 10^4$ GeV,
and $\alpha=-6.5 \times 10^{-3}\, {\rm GeV\,cm^2/g}$ and
$\beta = 3.66\times 10^{-6} \, {\rm cm^2/g}$ for higher energies \cite{KBS}.
Putting Eqs. (\ref{eq:esur}) and (\ref{eq:musur}) back into Eq. (\ref{eq:mudepth}) yields an expression for the muon flux as a function of energy and depth.  

\begin{figure}[h]
\includegraphics[width=2.3in,angle=270]{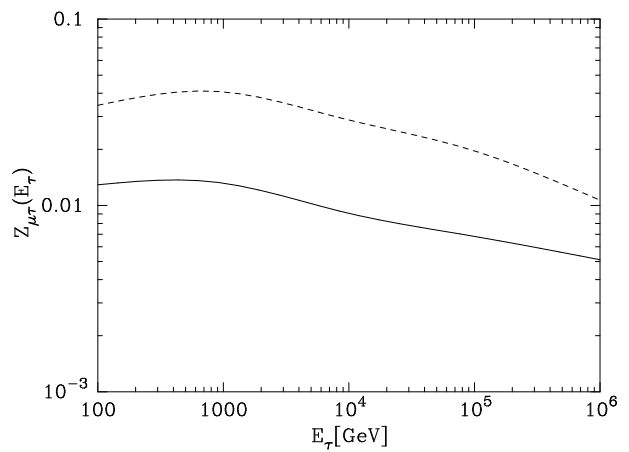}
\caption{Tau production $Z_{\mu\tau}$ moment for pair produced taus from incident muons.  Shown are the $Z_{\mu\tau}$ moments for the conventional (solid) and prompt (dashed) muon fluxes with the approximations of Eqs. (\ref{eq:mufac}) and (\ref{eq:mufacp}).}
\label{fig:ztau}
\end{figure}

An analytic solution to the cascade equations can be found when the fluxes can be factorized in the form $\phi_\mu(E,X)=\phi_1(E)\phi_2(X)$ \cite{lipari}.  For the tau neutrino energies of interest here, $E_{\nu_\tau}\geq 100$ GeV, $E_\mu\gg 100$ GeV, so the surface energy can be well approximated by $E_\mu^s\simeq E_\mu \exp(\beta X)$.  For the depth ranges of importance in our calculation, $\exp(-\beta X)\simeq 1$.  Using these approximations, the incident muon fluxes can be factorized according to 
\ba
\label{eq:mufac}
\phi_\mu^{\rm conv}(E_\mu,X) &\simeq &\phi_\mu^{\rm conv}(E_\mu,0)e^{-\beta\gamma X} \\ 
\label{eq:mufacp}
\phi_\mu^{\rm prompt}(E_\mu,X) &\simeq &\phi_\mu^{\rm prompt}(E_\mu,0)e^{-\beta\gamma' X}\ .
\ea

The muon fluxes go into the evaluation of $Z_{\mu\tau}$ in Eq. (\ref{eq:zmutau}).
Figure \ref{fig:ztau} shows the tau production $Z_{\mu\tau}$ moments using the
approximate muon flux in Eq. (\ref{eq:mufac}) for the conventional (solid line) flux and
Eq. (\ref{eq:mufacp}) for the prompt (dashed line) flux. The prompt $Z$-moment is larger than the conventional
$Z$-moment because the prompt muon flux has a smaller spectral index.

Using the factorized form of the muon flux at column depth $X$, the tau flux from
the conventional muon flux as a function of energy and depth is
\ba
\phi_\tau(E_\tau,X) &=& \frac{Z_{\mu\tau}(E_\tau)\phi_\mu(E_\tau,0)}{\lambda_{pair}(E_\tau)(1/\lambda_{dec}(E_\tau)-\beta\gamma)} \\ \nonumber
&\times& \Bigl[\exp(-\beta\gamma X)-\exp(-X/\lambda_{dec})\Bigr]\ .
\ea
A similar expression is obtained for the tau flux from prompt muons.
Because $\lambda_{dec}$ is a function of $E_\tau$,
there is not a factorizable form applicable to all energies, $\phi_\tau(E,X)\neq\phi_1(E)\phi_2(X)$ , so we have solved the cascade equation for
the tau neutrino flux generated by tau decays numerically. 

\section{Oscillated and prompt atmospheric tau neutrinos}

In addition to tau pair production and decay, there are two other sources of tau neutrinos. At low energies,
the main source of tau neutrinos is from oscillations of $\nu_\mu\to\nu_\tau$ \cite{pdg}.  In the absence of matter effects, 
a good approximation for the energy range of interest \cite{gandhi}, the probability of $\nu_\mu\to\nu_\tau$ can be expressed in terms of the atmospheric mixing angle $\theta_{\rm atm}$ and mass splitting $\Delta m^2$ in eV$^2$ in the two flavor approximation,
\be
\label{eq:prob}
P(\nu_\mu\to\nu_\tau) = \sin^2(2\theta_{\rm atm})\sin^2\Bigl(1.27\Delta m^2\frac{L}{E}\Bigr)\ ,
\ee
For a distance $L$ in km and energy $E$ in GeV.  For the atmospheric mixing angle and mass splitting, we use $\sin^2(2\theta_{\rm atm})=1$ 
and $\Delta m^2=2.5\times 10^{-3}\ {\rm eV}^2$ \cite{pdg}.  To calculate the tau neutrinos produced from oscillations of $\nu_\mu\to\nu_\tau$ for the downward atmospheric flux, one needs
to account for the production height. Gaisser and Stanev \cite{pathlength} have shown that for $E_\nu>100$ GeV, the
pathlength distribution (in $\ell$) of muon neutrinos is approximately flat
out to $\ell \cos\theta\simeq 20$ km, followed by a declining production rate. We use the Bartol flux of muon neutrinos at sea level $\phi_{\nu_\mu}\equiv\phi_{\nu_\mu+\bar{\nu}_\mu}(E_\nu,0)$ \cite{bartol}
with a distribution that is constant up to $L_{max}\cos\theta = 20$ km,
\be
\label{eq:phinumu}
\frac{d\phi_{\nu_\mu}(E_{\nu_\mu},\ell)}{d\ell}
 = \frac{\Theta (L_{max}-\ell)}{L_{max}}
 \phi_{\nu_\mu}\ .
\ee
In the high energy approximation  with our choice of mixing angle
and mass splitting, this gives a tau neutrino flux at the surface of the Earth of
\be
\phi_{\nu_\tau}\simeq \frac{1}{3}\times 10^{-5} 
\Bigl( \frac{L_{max}}{E}\Bigr)^2 \phi_{\nu_\mu} \ ,
\ee
where the factor of 1/3 accounts for the distribution in production
heights. For a detector at depth $d$, looking up at a zenith angle $\theta$,
the oscillated tau neutrino flux is approximately
\be
\phi_{\nu_\tau}\simeq\Biggl( \frac{L_{max}^2}{3}+\Bigl(\frac{d}{\cos\theta}\Bigr)^2\Biggr)
\times 10^{-5} 
\Bigl( \frac{1}{E}\Bigr)^2 \phi_{\nu_\mu} \ .
\ee
%We use $d=1.5$ km here.
For our results of downward neutrinos, we consider a depth $d=1.5$ km.

For upward neutrino fluxes, the height of production and
$d$ are negligible
compared to the length of the chord $D$ through the Earth that the
neutrinos travel, so we take the upward oscillated tau neutrino
flux to be
\be
\phi_{\nu_\tau}\simeq  10^{-5} 
\Bigl( \frac{D}{E}\Bigr)^2 \phi_{\nu_\mu}\ .
\ee

The other source of atmospheric tau neutrinos is the decay of $D_s$ mesons that are produced in cosmic ray interactions in the atmosphere \cite{Pasquali,stasto}.  An approximate form for the prompt flux of tau neutrinos and antineutrinos can be written \cite{stasto}
\begin{eqnarray}
\label{eq:nustasto}
\phi_{\nu_\tau+\bar{\nu}_\tau}(E,0) &=& \frac{1\times 10^{-7} E^{0.5}\ {\rm (GeV \, cm^2\, sr\, s)}^{-1}}{(E/{\rm GeV})^3} \\ \nonumber
&\times& \Bigl(\frac{1}{1+(E/1\times 10^6)^{0.7}+(E/4\times 10^6)^{1.5}}\Bigr) .
\end{eqnarray}  
At the energies of interest, the flux of prompt tau neutrinos is independent of zenith angle due to the short lifetime of the $D_s$ meson.

\begin{figure}[h]
\includegraphics[width=2.3in,angle=270]{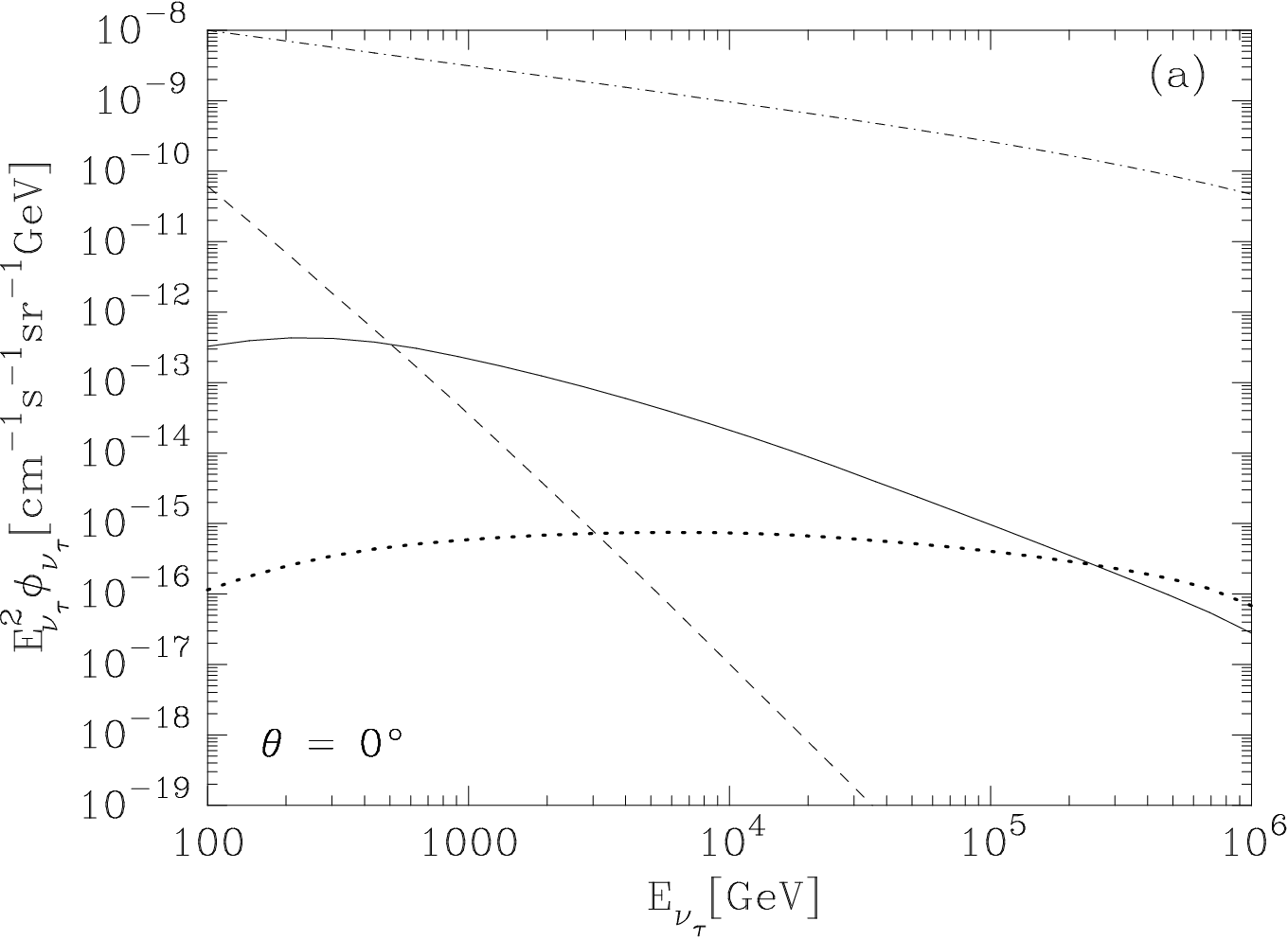}
\includegraphics[width=2.3in,angle=270]{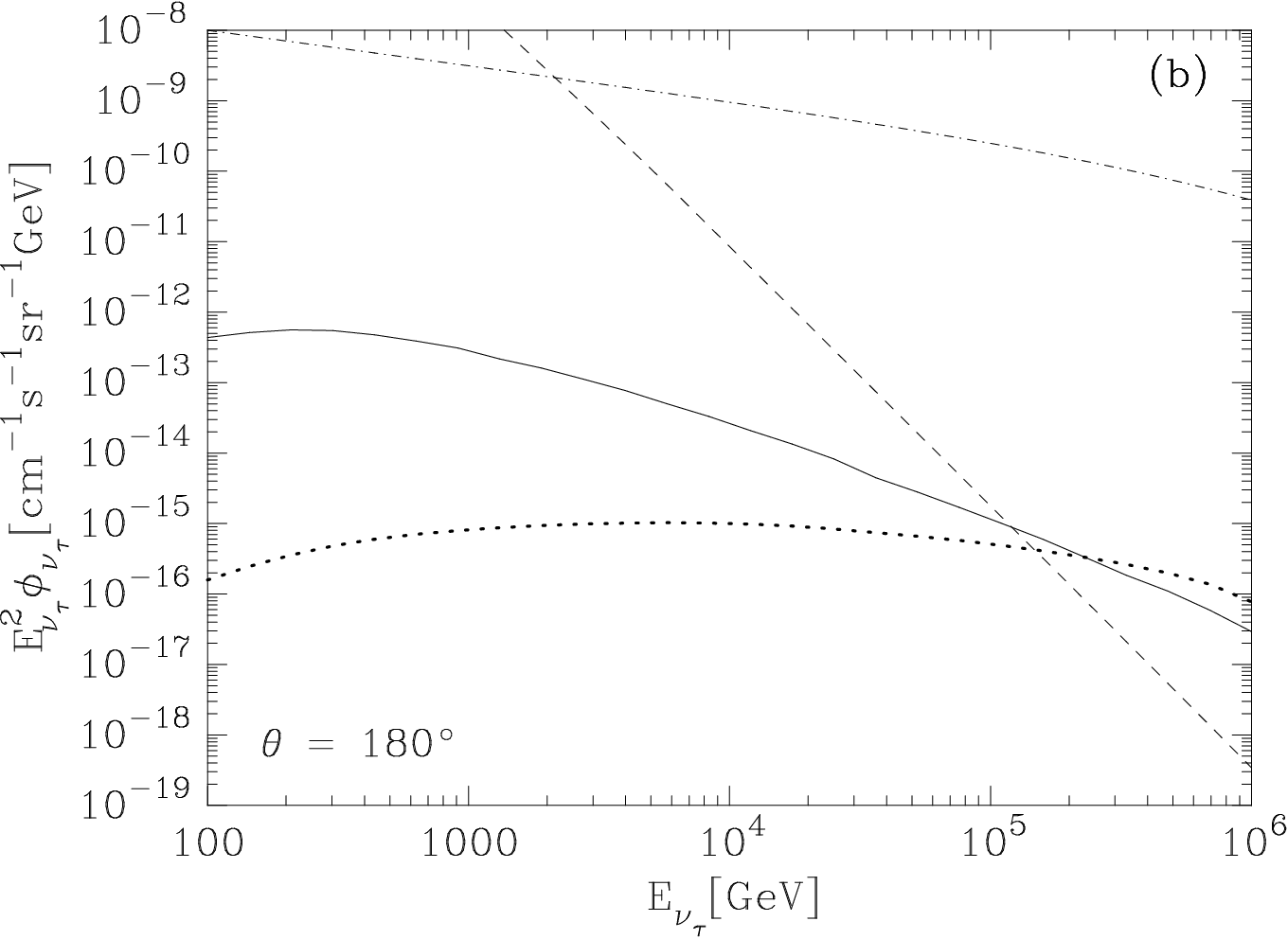}
\caption{Differential tau neutrino flux scaled by the square of the neutrino energy for neutrinos (a) going downward (with a zenith angle of 0$^\circ$)
and (b) going upward (with a zenith angle of 180$^\circ$).  
The solid curve represents the tau neutrino flux from the incident conventional
muon flux ($\mu\to\tau\to\nu_\tau$), and the dotted curve comes from the prompt atmospheric muons.
The dashed curve shows the oscillated conventional flux from muon neutrinos
and the dot-dashed curve shows the prompt tau neutrino flux itself.}
\label{fig:nudown}
\end{figure}

\section{Results and Discussion}

Our results for downward and upward tau neutrinos are shown
in Figs. 2(a) and 2(b). Figure 2(a) shows the $\mu\to\tau\to\nu_\tau$ contributions of
the conventional muon flux (solid line) and the prompt muon flux (dotted line),
in comparison to the oscillated conventional $\nu_\mu\to\nu_\tau$ flux for
downward neutrinos with a zenith angle $\theta=0^\circ$ (dashed line).
With the short distance over which muon neutrinos can oscillate to tau 
neutrinos, muon production of taus which then decay, dominate over the oscillated
conventional tau neutrino flux above $E_\nu\simeq 500$ GeV. 

Figure 2(b) shows the upward ($\theta = 180^\circ$) tau neutrino fluxes. There,
with the oscillation distance equal to the diameter of the Earth, the oscillated
flux dominates the  $\mu\to\tau\to\nu_\tau$ over a much larger energy range.  For upward neutrino fluxes, at the energies of interest, neutrino attenuation needs to be considered.  To account for neutrino attenuation of upward going neutrinos, we have included a factor of $\exp(-\sigma_{cc}XN_A/A)$, where $\sigma_{cc}$ is the tau neutrino charged current cross section.  For the charged current cross section, we have used the results from Ref. \cite{Jeong}.  This factor is a good approximation even for the $\mu\to\tau\to\nu_\tau$ flux because most of the tau pair production and subsequent decays happens near the surface of the earth.

The dot-dashed lines
in Figs. 2(a) and 2(b) shows the prompt $\nu_\tau+\bar{\nu}_\tau$ flux evaluated
in Ref. \cite{stasto} given by Eq. (\ref{eq:nustasto}) from
$D_s$ production followed by its decay to $\tau\nu_\tau$. While there is some
uncertainty in the prompt tau neutrino flux, the figures show that it dominates
the atmospheric contributions to the flux of $\nu_\tau+\bar{\nu}_\tau$ over a
wide range of energies discussed here.
The dominant contribution to the tau neutrino flux for
$E>2$ TeV \cite{leelin}, regardless of the zenith
angle direction, is the direct prompt production of tau neutrinos.
For downward neutrinos, the direct prompt production dominates the other
sources of tau neutrinos for energies larger than tens of GeV. 

In addition to downward ($\theta=0^\circ$) and upward 
($\theta=180^\circ$) tau neutrino fluxes, the flux of tau neutrinos at higher zenith angles can be calculated.  For higher zenith angles,
the conventional fluxes are increased, while the prompt fluxes are
essentially unchanged for $E<10^6$ GeV.
The increased conventional muon flux means that 
there is an increase in the flux of neutrinos produced via $\mu\to\tau\to\nu_\tau$.  Similarly, however, there is an increase in the flux of tau neutrinos from the longer pathlength for oscillations plus the higher flux of muon neutrinos.  At an incident zenith angle of $80^\circ$, tau neutrinos from $\mu\to\tau\to\nu_\tau$ dominate those from oscillations over the approximately the same energy range as for
$\theta=0^\circ$. At $\theta=80^\circ$, the conventional $\mu \to\tau\to\nu_\tau$ and $\nu_\mu\to\nu_\tau$ oscillated fluxes are
increased by a factor of about three, but the prompt $\nu_\tau$ flux
still dominates.  

In summary, while the flux of downward tau neutrinos produced from atmospheric muons dominates that produced by neutrino oscillations for a large range of energies and zenith angles, both of these sources are suppressed in comparison to the flux of prompt tau neutrinos coming from the decay of $D_s$ mesons.

\acknowledgments
This research was supported by the US Department of Energy Contract No.
DE=FG02-91ER40664. We thank I. Sarcevic for helpful conversations.

%=============================================================================


\begin{thebibliography}{99}
%=============================================================================
\bibitem{bartol}
 V.~Agrawal, T.~K.~Gaisser, P.~Lipari and T.~Stanev,
  %``Atmospheric neutrino flux above 1 GeV,''
  Phys.\ Rev.\  D {\bf 53}, 1314 (1996).
%  [arXiv:hep-ph/9509423].
  %%CITATION = PHRVA,D53,1314;%%
  
\bibitem{lipari}
See, e.g., P.~Lipari,
  %``Lepton spectra in the earth's atmosphere,''
  Astropart.\ Phys.\  {\bf 1}, 195 (1993).
  %%CITATION = APHYE,1,195;%%
  
\bibitem{pathlength}
  T.~K.~Gaisser and T.~Stanev,
  %``Path length distributions of atmospheric neutrinos,''
  Phys.\ Rev.\  D {\bf 57}, 1977 (1998).
%  [arXiv:astro-ph/9708146].
  %%CITATION = PHRVA,D57,1977;%%
  
\bibitem{prs}
L.~Pasquali, M.~H.~Reno and I.~Sarcevic,
  %``Lepton fluxes from atmospheric charm,''
  Phys.\ Rev.\  D {\bf 59}, 034020 (1999).
%  [arXiv:hep-ph/9806428].
  %%CITATION = PHRVA,D59,034020;%%
  
\bibitem{tig}
P.~Gondolo, G.~Ingelman and M.~Thunman,
  %``Charm production and high energy atmospheric muon and neutrino fluxes,''
  Astropart.\ Phys.\  {\bf 5}, 309 (1996).
 % [arXiv:hep-ph/9505417].
  %%CITATION = APHYE,5,309;%%

\bibitem{Pasquali}
  L.~Pasquali and M.~H.~Reno,
  %``Tau neutrino fluxes from atmospheric charm,''
  Phys.\ Rev.\  D {\bf 59}, 093003 (1999).
 % [arXiv:hep-ph/9811268].
  %%CITATION = PHRVA,D59,093003;%%

\bibitem{stasto}
A.~D.~Martin, M.~G.~Ryskin and A.~M.~Stasto,
  %``Prompt neutrinos from atmospheric $c \bar{c}$ and $b \bar{b}$ production
  %and the gluon at very small x,''
  Acta Phys.\ Polon.\  B {\bf 34}, 3273 (2003).
%  [arXiv:hep-ph/0302140].
  %%CITATION = APPOA,B34,3273;%%

\bibitem{superk}
  Y.~Fukuda {\it et al.}  [Super-Kamiokande Collaboration],
  %``Evidence for oscillation of atmospheric neutrinos,''
  Phys.\ Rev.\ Lett.\  {\bf 81}, 1562 (1998).
%  [arXiv:hep-ex/9807003].
  %%CITATION = PRLTA,81,1562;%%
  
\bibitem{drss}
 S.~I.~Dutta, M.~H.~Reno, I.~Sarcevic and D.~Seckel,
  %``Propagation of muons and taus at high energies,''
  Phys.\ Rev.\  D {\bf 63}, 094020 (2001).
  %[arXiv:hep-ph/0012350].
  %%CITATION = PHRVA,D63,094020;%%
    

\bibitem{br1}
A.~Bulmahn and M.~H.~Reno,
  %``Cross sections and energy loss for lepton pair production in muon
  %transport,''
  Phys.\ Rev.\  D {\bf 79}, 053008 (2009).
 % [arXiv:0812.5008 [hep-ph]].
  %%CITATION = PHRVA,D79,053008;%%
  

\bibitem{KBS}
  S.~I.~Klimushin, E.~V.~Bugaev and I.~A.~Sokalski,
  %``On the parametrization of atmospheric muon angular flux underwater,''
  Phys.\ Rev.\  D {\bf 64}, 014016 (2001).
 % [arXiv:hep-ph/0012032].
  %%CITATION = PHRVA,D64,014016;%% 

\bibitem{br2}
 A.~Bulmahn and M.~H.~Reno,
  %``High energy leptons from muons in transit,''
  Phys.\ Rev.\  D {\bf 81}, 053003 (2010).
%  [arXiv:0912.1385 [hep-ph]].
  %%CITATION = PHRVA,D81,053003;%%

\bibitem{enberg}
 R.~Enberg, M.~H.~Reno and I.~Sarcevic,
  %``Prompt neutrino fluxes from atmospheric charm,''
  Phys.\ Rev.\  D {\bf 78}, 043005 (2008).
%  [arXiv:0806.0418 [hep-ph]].
  %%CITATION = PHRVA,D78,043005;%%
  
\bibitem{pdg}
C. Amsler et al. (Particle Data Group), Phys. Lett. {\bf B 667},
1 (2008).

\bibitem{gandhi}
R.~Gandhi, P.~Ghoshal, S.~Goswami, P.~Mehta and S.~Uma Sankar,
  %``Earth matter effects at very long baselines and the neutrino mass
  %hierarchy,''
  Phys.\ Rev.\  D {\bf 73}, 053001 (2006).
%  [arXiv:hep-ph/0411252].
  %%CITATION = PHRVA,D73,053001;%%

\bibitem{leelin}
F.~F.~Lee and G.~L.~Lin,
  %``A Semi-analytic calculation on the atmospheric tau neutrino flux in the GeV
  %to TeV range,''
  Astropart.\ Phys.\  {\bf 25}, 64 (2006).
%  [arXiv:hep-ph/0412383].
  %%CITATION = APHYE,25,64;%%

\bibitem{Jeong}
  Y.~S.~Jeong and M.~H.~Reno,
  %``Tau neutrino and antineutrino cross sections,''
  arXiv:1007.1966 [hep-ph].
  %%CITATION = ARXIV:1007.1966;%%

 

\end{thebibliography}
\end{document}